\documentclass[%
 reprint,
 superscriptaddress,
 amsmath,amssymb,
 aps,
 prx,
]{revtex4-2}

\usepackage{graphicx}
\usepackage{amsmath}
\usepackage{mathtools}
\usepackage{comment}
\usepackage{bbm}
\usepackage{lipsum}
\usepackage{amssymb}
\usepackage{xcolor}
\definecolor{red}{rgb}{1,0.,0}

\usepackage{graphicx}% Include figure files
\usepackage{subcaption}
\usepackage{dcolumn}% Align table columns on decimal point
\usepackage{bm}% bold math

\begin{document}

%\preprint{APS/123-QED}

\title{Multifunctional Superconducting Nanowire Quantum Sensors}% Force line breaks with \\

\author{Benjamin J. Lawrie}
\email{lawriebj@ornl.gov; This manuscript has been authored by UT-Battelle, LLC, under contract DE-AC05-00OR22725 with the US Department of Energy (DOE). The US government retains and the publisher, by accepting the article for publication, acknowledges that the US government retains a nonexclusive, paid-up, irrevocable, worldwide license to publish or reproduce the published form of this manuscript, or allow others to do so, for US government purposes. DOE will provide public access to these results of federally sponsored research in accordance with the DOE Public Access Plan (http://energy.gov/downloads/doe-public-access-plan). }
\address{Materials Science and Technology Division, Oak Ridge National Laboratory, 1 Bethel Valley Rd, Oak Ridge, TN 37831} 
\author{Claire E. Marvinney}
\email{marvinneyce@ornl.gov}
\address{Materials Science and Technology Division, Oak Ridge National Laboratory, 1 Bethel Valley Rd, Oak Ridge, TN 37831} 
\author{Yun-Yi Pai}
\address{Materials Science and Technology Division, Oak Ridge National Laboratory, 1 Bethel Valley Rd, Oak Ridge, TN 37831} 
\author{Matthew A. Feldman}
\address{Materials Science and Technology Division, Oak Ridge National Laboratory, 1 Bethel Valley Rd, Oak Ridge, TN 37831} 
\author{Jie Zhang}
\address{Materials Science and Technology Division, Oak Ridge National Laboratory, 1 Bethel Valley Rd, Oak Ridge, TN 37831} 
\author{Aaron J. Miller}
\address{Quantum Opus LLC, 22241 Roethel Dr Ste A, Novi, MI 48375}
\author{Chengyun Hua}
\address{Materials Science and Technology Division, Oak Ridge National Laboratory, 1 Bethel Valley Rd, Oak Ridge, TN 37831} 
\author{Eugene Dumitrescu}
\address{Computational Science and Engineering Division, Oak Ridge National Laboratory, 1 Bethel Valley Rd, Oak Ridge, TN 37831} 
\author{G\'abor B. Hal\'asz}
\address{Materials Science and Technology Division, Oak Ridge National Laboratory, 1 Bethel Valley Rd, Oak Ridge, TN 37831}

\date{\today}% It is always \today, today,

\begin{abstract}
Superconducting nanowire single photon detectors (SNSPDs) offer high-quantum-efficiency and low-dark-count-rate single photon detection. In a growing number of cases, large magnetic fields are being incorporated into quantum microscopes, nanophotonic devices, and sensors for nuclear and high-energy physics that rely on SNSPDs, but superconducting devices generally operate poorly in large magnetic fields. Here, we demonstrate robust performance of amorphous SNSPDs in magnetic fields of up to $\pm 6$ T with a negligible dark count rate and unchanged quantum efficiency at typical bias currents. Critically, we also show that in the electrothermal oscillation regime, the SNSPD can be used as a magnetometer with sensitivity of better than 100 $\mathrm{\mu T/\sqrt{Hz}}$ and as a thermometer with sensitivity of 20 $\mathrm{\mu K/\sqrt{Hz}}$ at 1 K. Thus, a single photon detector integrated into a quantum device can be used as a multifunctional quantum sensor capable of describing the temperature and magnetic field on-chip simply by varying the bias current to change the operating modality from single photon detection to thermometry or magnetometry.

\end{abstract}

\pacs{Valid PACS appear here}

\maketitle

Superconducting nanowire single photon detectors (SNSPDs) offer high speed, high quantum-efficiency, and low dark-count-rate single photon detection~\cite{natarajan2012superconducting}.  There is an emerging need for SNSPDs capable of operating in large magnetic fields for potential integration with quantum nanophotonic circuits ~\cite{khasminskaya2016fully,dreyer2018first,rath2015superconducting} and for quantum sensors relevant to nuclear physics ~\cite{azzouz2012efficient} and dark matter detection ~\cite{berggrensnowmass2021,jplsnowmass2021,hochberg2019detecting}. In general, SNSPDs do not perform well in large magnetic fields, but some research has demonstrated that small magnetic fields can improve SNSPD performance as described below.

SNSPDs can detect photons when (a) the photon provides enough energy to break an ensemble of Cooper pairs and generate a bath of quasiparticles that form a belt across the width of the nanowire, (b) the photon provides enough energy to unpin a vortex, enabling it to sweep across the nanowire under a Lorentz force, and (c) the photon provides enough energy to form and unbind a vortex-antivortex pair that are swept in opposite directions across the nanowire under a Lorentz force. Field-dependent studies of bright count rates have suggested that vortex motion is the primary detection mechanism in typical NbN SNSPDs~\cite{vodolazov2015vortex,lusche2014effect}. Similar measurements of $\mathrm{Mo_xSi_{1-x}}$ SNSPDs with varying wire width have concluded that vortex-antivortex interactions are responsible for bright counts in wide devices made from micron-scale wires while direct quasiparticle belt formation is responsible for bright counts in nanowires of order 100 nm in width~\cite{korneeva2020different}. A growing body of literature has made it clear that SNSPDs constructed of different materials with different designs can rely on any of the bright-count mechanisms described above~\cite{engel2014detection}. 

Dark counts are observed as a result of thermally induced vortex motion in the absence of any photons, and are generally present even in well shielded detectors operated at 0 T~\cite{bulaevskii2011vortex,bulaevskii2012vortex}. Because the local vortex pinning potential can be a spatially heterogeneous function of the superconducting film growth conditions and the device geometry, dark counts may arise preferentially from weak spots within the device. Early models suggested that the energy barrier for single vortex crossing is lower than that for phase slips and vortex-antivortex nucleation and annihilation~\cite{bulaevskii2011vortex} and that vortex-antivortex interactions and phase slips may be ignored, but competing models and measurements have suggested that vortex-antivortex interactions resulting from a Berezinskii-Kosterlitz-Thouless transition are responsible for SNSPD dark counts~\cite{yamashita2011origin,engel2006fluctuation,kitaygorsky2007dark}. 

At high bias currents, hotspots periodically form and disappear with frequency determined by the device inductance and load impedance, leading to detection of dark counts at this electrothermal oscillation frequency ~\cite{hadfield2005low,kerman2009electrothermal,liu2013electrical,berggren2018superconducting,marvinney2021waveform}. The propagation velocity of a normal-superconducting boundary in the nanowire due to Joule heating depends on its critical current \cite{kerman2009electrothermal}, which itself has a strong dependence on the applied external magnetic field. While detection events in the electrothermal oscillation regime are a result of a fundamentally different physical process than dark counts detected at lower bias currents, we describe both conventional dark counts and electrothermal oscillations as dark counts in this article for linguistic convenience.

Current crowding at sharp bends can result in a reduced potential barrier for vortex motion and an increased vortex nucleation density. However, external magnetic fields and associated Meissner currents can reduce the effect of current crowding~\cite{clem2012predicted,ilin2014magnetic}. Early modeling efforts suggested that perpendicular magnetic fields would reduce the critical current for a conventional meander line SNSPD, but small negative perpendicular magnetic fields could increase the critical current of devices patterned in a spiral layout~\cite{clem2012predicted}. Other experimental efforts found a slight asymmetry in the dark count rate of TaN and NbN SNSPDs as a function of field~\cite{engel2012dependence,charaev2019magnetic}, consistent with previous modeling~\cite{clem2012predicted}, subject to the assumption that dark counts originate largely at the sharp corners of the device and allowing for some heterogeneity in the device fabrication~\cite{engel2012dependence}. 

Understanding and controlling vortex motion and hotspot formation in SNSPDs is critical to the development of quantum sensors capable of operating in high magnetic fields, but fully predictive models of single-photon interactions with superconducting nanowires remain a challenge, despite the recent demonstration of a probabilistic criterion for single photon detection based on single vortex motion~\cite{jahani2020probabilistic}. Very little work has explored the field dependence of SNSPDs in large magnetic fields, though one recent article did demonstrate that NbN nanowire single photon detectors can be operated in fields as high as 5 T~\cite{polakovic2020superconducting}, and many researchers have explored the response of superconducting thin films and nanowires in larger magnetic fields~\cite{gardner2011enhancement,PhysRevB.32.2190,rouco2019depairing,cordoba2013magnetic,samkharadze2016high,zhang2016characteristics}. Further research is needed to provide an improved understanding of vortex motion in superconducting nanowires and to define the limits of high-field SNSPD operation.

\begin{figure}[b]
\centering
    \includegraphics[width=\columnwidth]{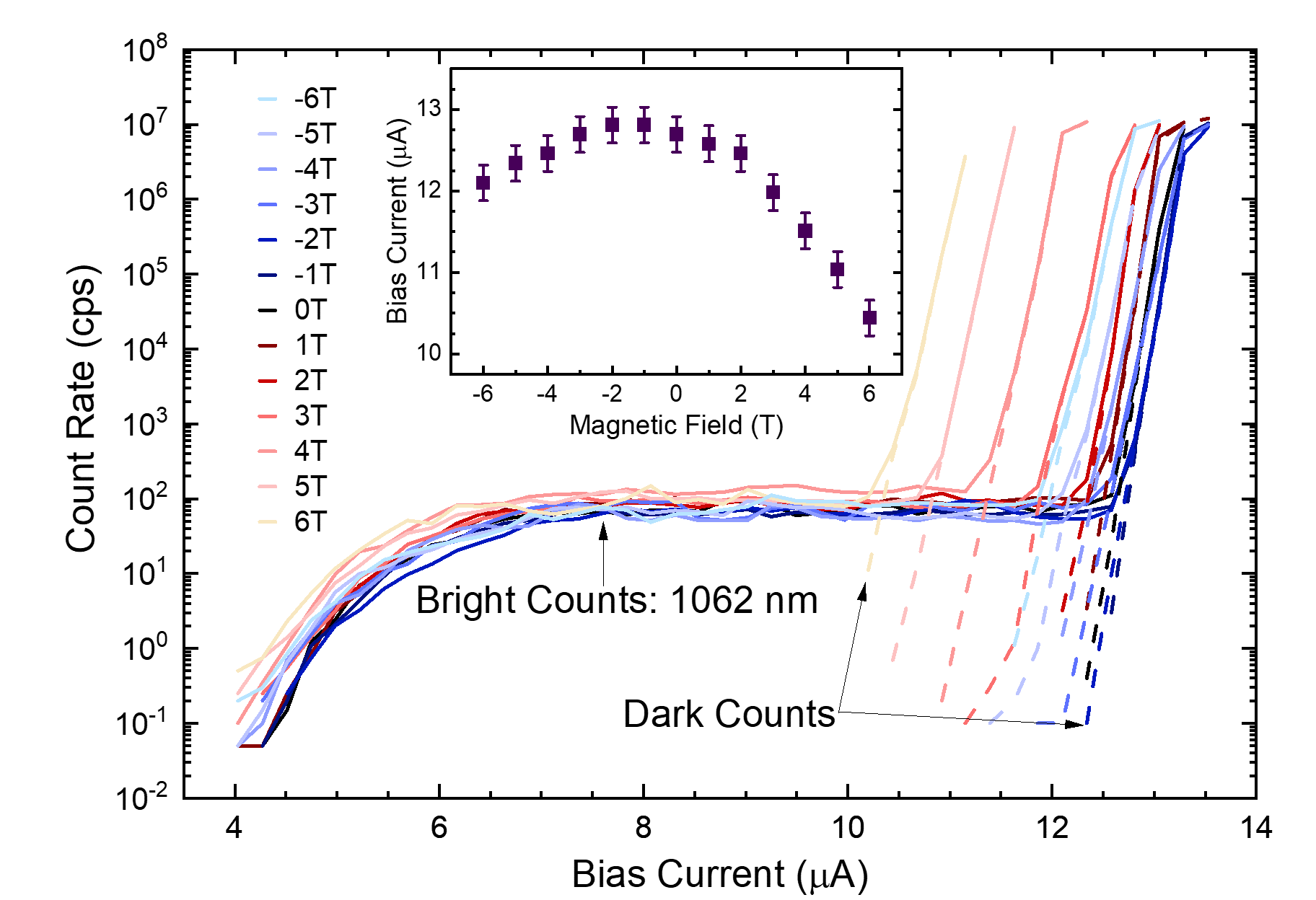}
    \caption{Measured count rate as a function of bias current and magnetic field for bias currents of 0-14 $\mu$A and magnetic fields of -6 to 6 T oriented roughly parallel to the film and $\sim 45^{\circ}$ to the length of the nanowires. Dark counts begin to contaminate the bright count measurements at bias currents above 10.0 $\mu$A. Inset of the bias current at which the dark counts are equal to 10x the plateau bright counts, emphasizing the asymmetry of the field dependence of dark counts at high bias currents.
}
    \label{fig:fig1}
\end{figure}

Here, we explore the magnetic field dependence of dark counts and bright counts generated by a commercially available amorphous transition-metal silicide near-infrared SNSPD with a critical temperature of 5 K. In order to minimize the density of vortices generated by the magnetic field, we focus on magnetic fields oriented parallel to the device. The 7 nm thick SNSPD used here has a 70 nm wide meander line with a 50 percent fill fraction that spans an 11 x 11 $\mu$m active area (Quantum Opus, LLC).  It was mounted on a PCB and suspended in a dilution refrigerator with free-space optical access to the mixing chamber at an angle of $<5^{\circ}$ from a magnetic field that was swept from -6 to 6 T. The SNSPD meander line pattern was oriented at an angle of $\sim 45^{\circ}$ relative to the field. Unless otherwise specified, the SNSPD temperature was held at 100 mK. An attenuated 1062 nm CW laser source was delivered to the SNSPD through a free-space optical interface with the dilution refrigerator that is described elsewhere~\cite{rsi}. Dark counts were collected with optical access to the dilution refrigerator blocked. All signals were passed through a low frequency filter mounted directly to the SNSPD at the mixing chamber in order to minimize latching and enable high count rate detection \cite{miller2020noise}.  The bias current and the SNSPD waveforms were delivered along the same semirigid coaxial channel with a room temperature bias tee and low noise amplifiers integrated before time tagging all detection events with a 62 mV threshold. Data was collected (1) continuously for counts as a function of magnetic field for a constant bias current and (2) in 1 T increments for counts as a function of bias current.  

\begin{figure}[b]
\centering
    \includegraphics[width=\columnwidth]{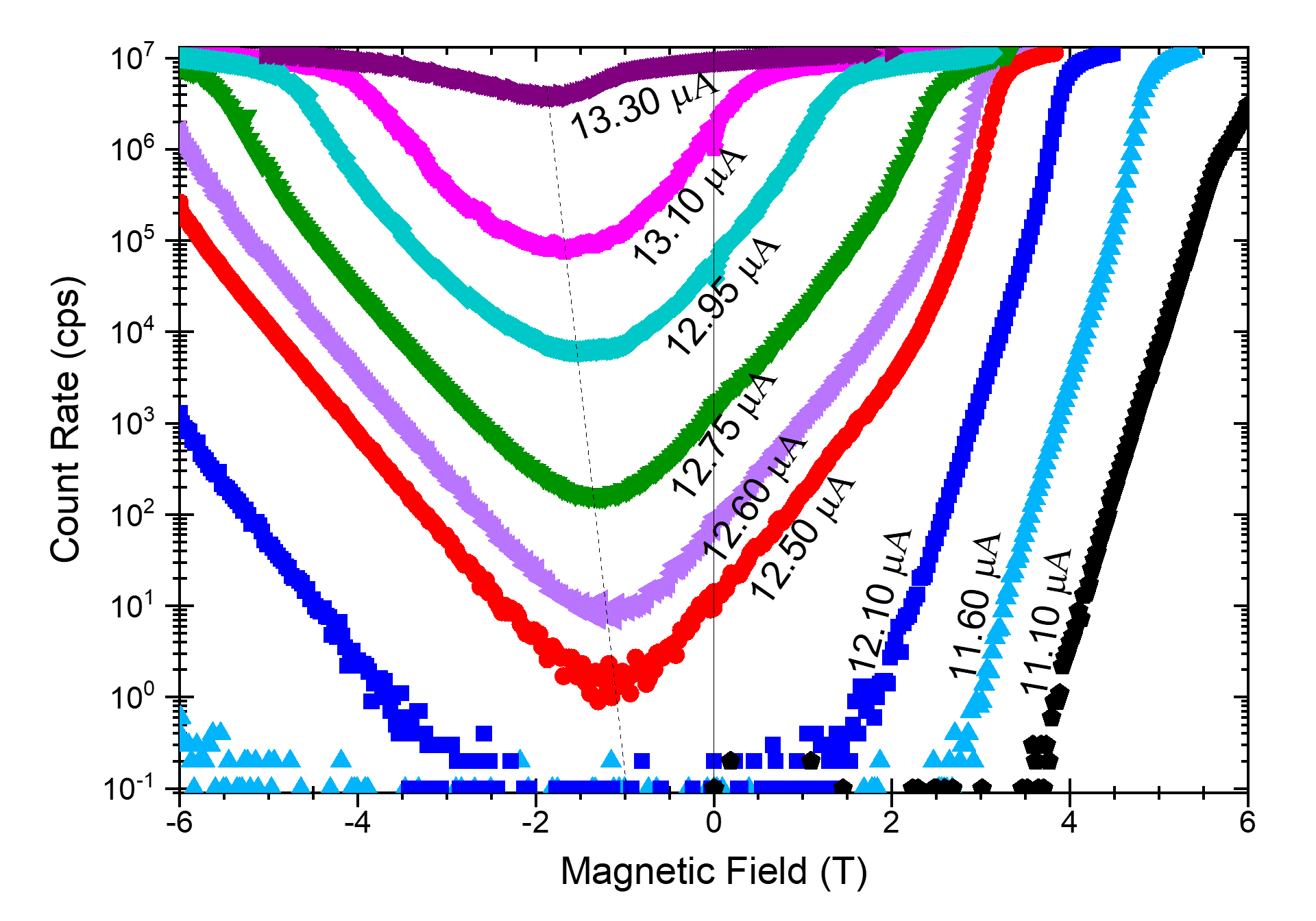}
    \caption{Dark count rate as a function of field for bias currents of 11.10-13.30 $\mu$A.
}
    \label{fig:fig2}
\end{figure}

The magnetic field dependence of the measured counts is shown as a function of bias current in Fig.~\ref{fig:fig1}.  At each point on the curve, the bright count rate was averaged for 20 seconds and the dark count rate was averaged for 10 seconds.  At all field values, the relative quantum efficiency of the device remains unchanged at bias currents within the quantum efficiency plateau between 7.5-9.5 $\mu$A. Additionally, the dark count rate was measured to be $<0.1$ counts per second within the same 7.5-9.5 $\mu$A range for all measured fields.   
However, as emphasized in the inset of Fig.~\ref{fig:fig1}, the maximum operating bias current is suppressed in an asymmetric manner with increasing field, with positive parallel magnetic fields reducing the maximum operational bias current to $\sim$ 10.0 $\mu$A at 6 T, while the maximum operational bias current at -6 T is only reduced to $\sim$ 11.5 $\mu$A.  A similar, if less pronounced asymmetry is present in the onset of bright count detection events at bias currents of 4-5.5 $\mu$A as seen in Fig.~\ref{fig:fig1}.

This asymmetric magnetic field response leads to a decrease in the dark count rate of the device at some negative fields, and thus an improved operation of the device between 0 T and -2.5 T, as shown in Fig.~\ref{fig:fig2}. This asymmetry was repeatable for multiple zero-field recooling measurements.  The transport properties of the high-count-rate board were also tested separately as a function of field, resulting in no evidence of asymmetry.  Thus the asymmetric response shown in Fig.~\ref{fig:fig2} appears to be due to the SNSPD itself.  While slight asymmetries in the field-dependent SNSPD response have been reported before for mT-scale perpendicular fields and ascribed to magnetic-field induced currents that oppose current crowding effects at the corners of the meander line~\cite{engel2012dependence,charaev2019magnetic,clem2012predicted}, the asymmetries reported here rely on similar bias currents but magnetic fields three orders of magnitude larger. Thus, the previous models for asymmetric SNSPD responses to magnetic fields do not appear to explain this result. Notably, while the onset of bright counts and dark counts is a function of magnetic field, and while the SNSPD waveform can be a function of the photon-nanowire interaction parameters~\cite{marvinney2021waveform,cahall2017multi}, the SNSPD waveforms recorded here were independent of field. 

While a detailed model of this field-reversal asymmetry is beyond the scope of this article, we can understand why the asymmetry is much larger than in previous works by recognizing the different orientation of the magnetic field and considering the symmetries of the meander line and the bias current. The amorphous superconductor is deposited onto a 120 nm $\mathrm{SiO_2}$ backreflector and the meander line is then capped with a 1.5 nm conformal $\mathrm{Al_2O_3}$ coating to prevent oxidation before 100 nm $\mathrm{SiO_2}$ and $\mathrm{Si_3N_4}$ antireflective coating layers are deposited.  If the top and bottom interfaces of the SNSPD are not identical, as is the case here, a field with a finite component along the long line segments is expected to have a field-reversal asymmetry. Indeed, based on the model in Ref.~\cite{clem2012predicted}, such a field should, depending on its sign, induce current crowding close to the top or the bottom interface, thus giving different rates for positive and negative fields. Moreover, even if the two interfaces are identical, a field-reversal asymmetry is expected if all three components of the field (parallel to the long line segments, the short line segments, and the perpendicular direction, respectively) are finite. In contrast, such a field-reversal asymmetry is forbidden for a purely perpendicular field, regardless of whether the top and bottom interfaces are identical or distinct. For a perpendicular field, it can then only appear due to imperfections of the meander line~\cite{engel2012dependence} and is expected to be much smaller.

\begin{figure}[t]
\centering
    \includegraphics[width=\columnwidth]{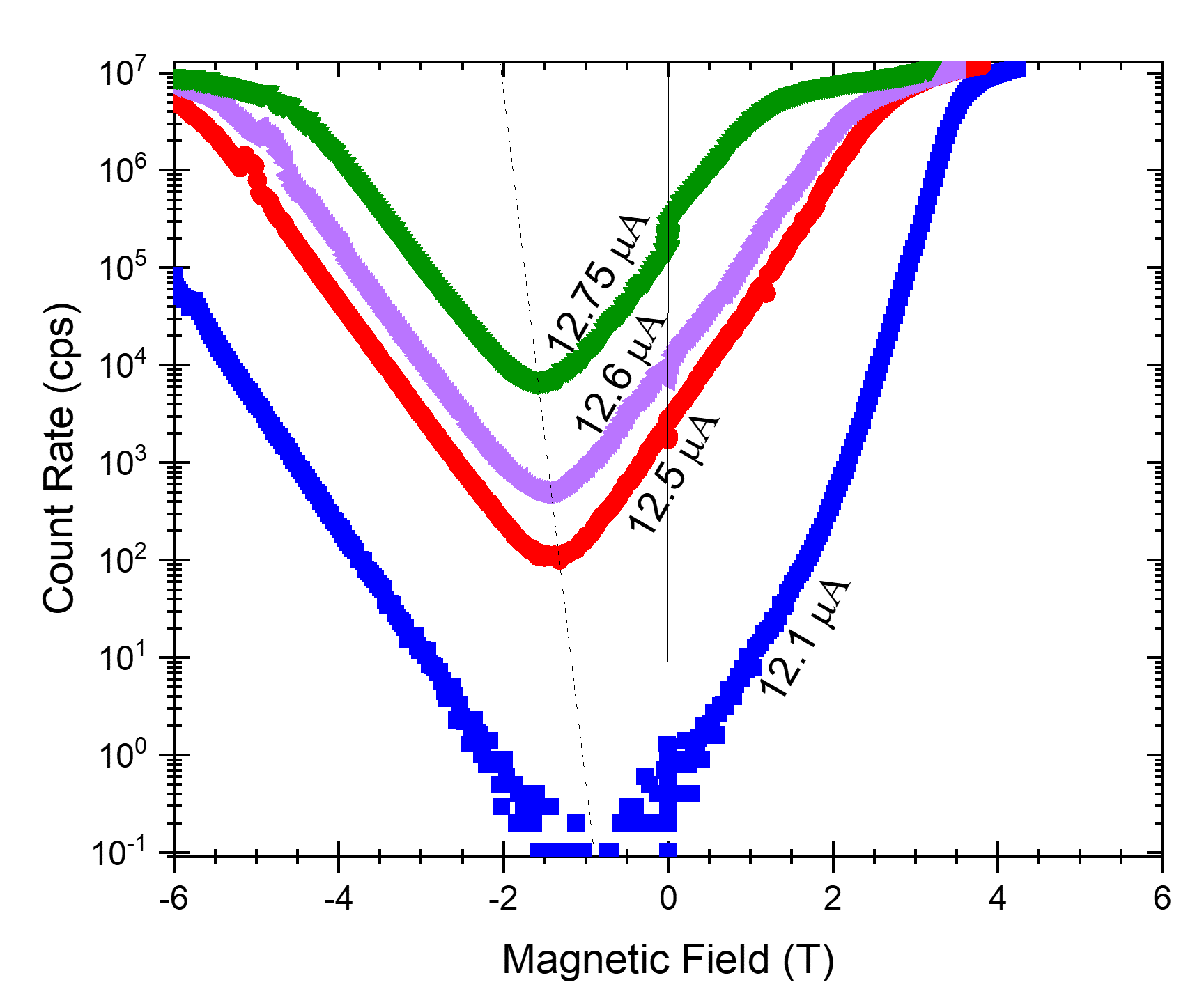}
    \caption{Dark count rate as a function of field for bias currents of 12.10-12.75 $\mu$A at $\sim$600 mK.
}
    \label{fig:fig3}
\end{figure}

The magnetic-field dependence of the dark counts in the high-bias-current regime is intriguing because it suggests that the SNSPD can be utilized as an on-chip magnetometer.  For a well characterized SNSPD, monitoring changes in the dark count rate after initially setting the bias current to achieve dark counts of $\sim 10^2-10^6 cps$ provides magnetic field sensitivity determined by the slope of the curves shown in Figs~\ref{fig:fig2} and \ref{fig:fig3} and limited by the uncertainty in counting statistics and bias current. Here, we assume that the field sensitivity can be calculated based on exponential fits to the measured dark counts and that the measured dark counts exhibit Poissonian counting statistics with uncertainty in count rate scaling as $\sqrt{N}$. Uncertainty in the bias current is neglected as a small component relative to the uncertainty in count rate. Further, uncertainty in photon counting is a fundamental limit that cannot be improved upon for this type of measurement, whereas the uncertainty in bias current can be improved beyond current limits with further engineering. 

The maximum sensitivity of the curves shown in Fig.~\ref{fig:fig2} ranges from 1.8 $mT/\sqrt{Hz}$ for a bias current of 13.3 $\mu$A at -1 T to 600 $\mu T/\sqrt{Hz}$ for a bias current of 11.10 $\mu$A at 6T, with the best sensitivity at smaller bias currents closer to the onset of the electrothermal oscillation regime.  The best sensitivity measured here was 75 $\mu T/\sqrt{Hz}$ for a bias current of 11.60 $\mu A$ at 4.8 T. Thus, the SNSPD can be used as a magnetometer by sweeping the bias current while monitoring the dark count rate to coarsely determine the magnetic field and by monitoring changes in the dark count rate at a constant bias current to track smaller changes in the magnetic field.

Extrapolating the results shown in Fig.~\ref{fig:fig1} to larger fields suggests that a plateau with optimized quantum efficiency and minimized dark counts should still exist for fields as large as 18 T and -68 T. It is likely that this extrapolation overestimates the robustness of the SNSPD, as the SNSPD operation relies in part on the description of the SNSPD as a 2D superconductor.  For fields above 6.5 T, the characteristic length scale $\sqrt{\hbar/2eB}$ is smaller than the 7 nm film thickness, and that description fails~\cite{PhysRevB.32.2190}.  Nonetheless, it is clear from the results in Fig.~\ref{fig:fig1} that the SNSPD is capable of robust operation for fields much larger than 6 T.

At higher temperatures, the maximum functional bias current is reduced as thermally induced vortex motion begins to increase, but the same robust high quantum-efficiency, low dark-count-rate operation is observed at all measured fields for bias currents of 6-12.1 $\mu$A. As seen in Fig.~\ref{fig:fig3} for dark counts measured at $\sim 600$ mK, a similar asymmetry is present in the dark count rate, with the minimum dark counts measured for fields between -1 T and -2 T.

\begin{figure}[t]
\centering
    \includegraphics[width=\columnwidth]{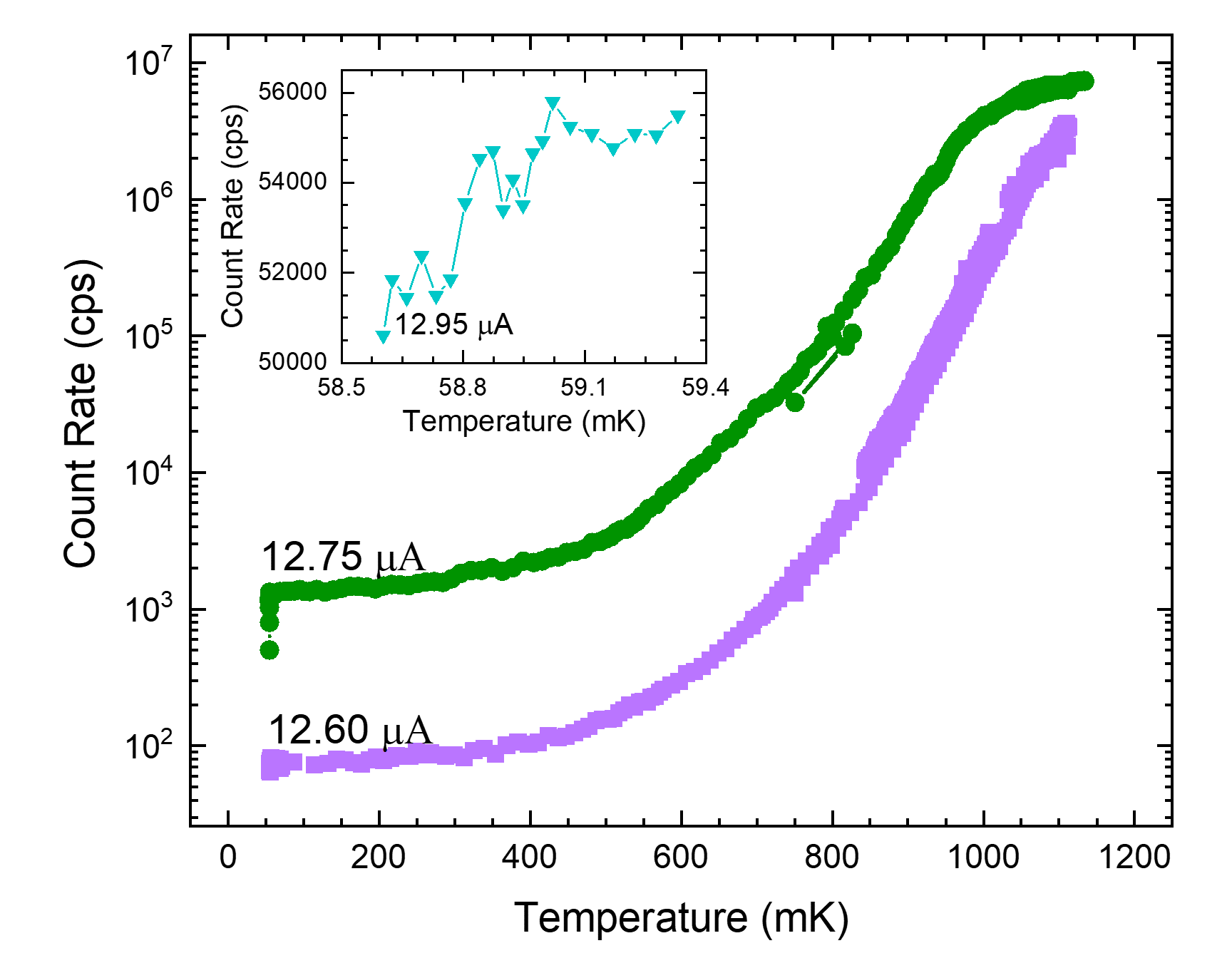}
    \caption{Temperature dependent dark counts for 12.60 $\mu A$ and 12.75 $\mu A$ bias currents at 0 T. The inset illustrates the temperature sensitivity at lower temperatures for a 12.95 $\mu A$ bias current at 0T.
}
    \label{fig:fig4}
\end{figure}

Because this SNSPD was mounted to a PCB at a distance of several inches from the nearest thermometer, there is some uncertainty in the SNSPD temperature that would normally be challenging to quantify. However, the dark counts can be used for thermometry just as they can be used for magnetometry. Figure \ref{fig:fig4} illustrates the measured dark counts for  12.60 $\mu A$ and 12.75 $\mu A$ bias currents at 0 T.  As above, the sensitivity of this superconducting nanowire thermometer can be calculated by curve fitting the dark count rate and assuming that the uncertainty in the count rate is limited by the photon counting statistics. The data shown in Fig.~\ref{fig:fig4}a illustrate temperature sensitivity of 20 $\mu K/\sqrt{Hz}$ at a temperature of 1 K and a bias of 12.75 $\mu A$, and a temperature sensitivity of 45 $\mu K/\sqrt{Hz}$ for a 12.60 $\mu A$ current.  At 800 mK, the sensitivity is reduced to 0.1 $mK/\sqrt{Hz}$ and 1.5 $mK/\sqrt{Hz}$, and at 400 mK, the sensitivity is reduced to 6 $mK/\sqrt{Hz}$ and 40 $mK/\sqrt{Hz}$, respectively at the larger and smaller bias currents. However, further increasing the bias current does continue to increase the sensitivity, and the SNSPD thermometer is most sensitive when the device is operating in the electrothermal oscillation regime with greater than $10^6$ dark counts per second.  With the device operating at a higher bias current, this regime can be reached at lower temperatures, and the onset of the electrothermal oscillations near 100 mK will improve the temperature sensitivity.  For example, on the cusp of the electrothermal oscillating regime at 60 mK, the sensitivity is improved to better than 0.5 $mK/\sqrt{Hz}$, as shown with a bias current of 12.95 $\mu A$ in Fig.~\ref{fig:fig4}b.  This improved sensitivity comes at the expense of increasing the device temperature $\sim 5$ mK at a bias current of 14 $\mu A$.

The ability to use the SNSPD in large magnetic fields for thermometry or magnetometry in addition to single photon detection is of interest, not just to better quantify the SNSPD operating conditions, but to provide a multifunctional sensor in integrated quantum nanophotonic devices. While the results described in this article provide a device-level understanding of the limitations of thermometry and magnetometry with SNSPDs in large fields, a microscopic understanding of the quasiparticle and vortex interactions that drive this functionality is still needed. Confocal optical microscopies capable of monitoring the SNSPD response as a function of the position and wavelength of incident photons at milliKelvin temperatures and in variable magnetic fields would provide an improved understanding of these interactions at the mesoscale. Additionally, a quantitative model of the magnetic field dependent and temperature dependent electrothermal oscillation frequencies will be pursued as a future work. MilliKelvin scanning probe microscopies including scanning SQUID microscopies~\cite{ceccarelli2019imaging} and scanning single photon microscopies~\cite{rsi} could help to provide further understanding of these interactions.

\acknowledgments
This research was sponsored by the U. S. Department of Energy, Office of Science, Basic Energy Sciences, Materials Sciences and Engineering Division. Postdoctoral (CEM) research support was provided by the Intelligence Community Postdoctoral Research Fellowship Program at the Oak Ridge National Laboratory, administered by Oak Ridge Institute for Science and Education through an interagency agreement between the U.S. Department of Energy and the Office of the Director of National Intelligence. Student (MAF, BEL) research support were provided by the Department of Defense through the National Defense Science $\&$ Engineering Graduate Fellowship (NDSEG) and by the DOE Science Undergraduate Laboratory Internships (SULI) program.

%\bibliography{references}

%3. unbinding of virtual vortex-antivortex pairs mimics Berezinskii-Kosterlitz-Thouless behavior and is tunable with film geometry\cite{gurevich2008size} 4. vortex core energy and realistic boundary conditions are necessary to model thermally activated hopping of vortices across superconducting films and explain discrepancies between (1,2) and (3)\cite{gurevich2012comment}.In WSi, photons can generate large normal-conducting areas without a bias current, yielding a different photon detection mechanism than in NbN. 

\end{document}